\documentclass[letterpaper]{article}

\usepackage{aaai}
\usepackage{times}
\usepackage{helvet}
\usepackage{courier}
\usepackage{graphicx}
\usepackage{url}

\usepackage{amsmath}
\usepackage{algorithm}
\usepackage[noend]{algpseudocode}
\makeatletter
\def\BState{\State\hskip-\ALG@thistlm}
\makeatother

\algnewcommand\algorithmicforeach{\textbf{for each}}
\algdef{S}[FOR]{ForEach}[1]{\algorithmicforeach\ #1\ \algorithmicdo}

\newcommand{\turkey}{{\em Mmm}Turkey}

\usepackage{xcolor}

\usepackage{soul}  
\setstcolor{red} 

\usepackage[textsize=tiny]{todonotes}
\begin{document}
%
\title{\turkey{}: A Crowdsourcing Framework for Deploying Tasks and 
Recording Worker Behavior on Amazon Mechanical Turk}

\author{Brandon Dang \\
Department of Computer Science \\
The University of Texas at Austin \\
budang@utexas.edu
\And Miles Hutson \\
Department of Computer Science \\
The University of Texas at Austin \\
mileshutson@utexas.edu
%
\And Matthew Lease \\
School of Information \\
The University of Texas at Austin \\
ml@utexas.edu
}

\maketitle

\begin{abstract}
\begin{quote}
Internal HITs on Mechanical Turk can be programmatically restrictive, and as a result, many requesters turn to using external HITs as a more flexible alternative. However, creating such HITs can be redundant and time-consuming.
We present 
\turkey{}\footnote{Not to be confused with {\em MmTurkey}, \url{https://github.com/longouyang/mmturkey}}, 
a framework that enables researchers to not only quickly create and manage external HITs, but more significantly also capture and record detailed worker behavioral data characterizing how each worker completes a given task.
\end{quote}
\end{abstract}

\section{Introduction}
\label{sec:intro}
Mechanical Turk 
{\em requesters} post HITs (Human Intelligence Tasks) to be completed by \textit{workers}. However, a challenge constantly faced by requesters is preserving the integrity of the data they collect. While most workers undertake tasks in good faith and are competent to deliver quality work, not all submitted work is high quality. This poses a potential risk to corrupt a data set \cite{eickhoff2013increasing}. Additionally, reported automated "botters" performing tasks meant for human workers further risk compromising data quality \cite{difallah2012mechanical}.

Just as search engines routinely capture user interactions in order to better understand their users and deliver higher quality results, prior work has suggested instrumenting worker interfaces can provide similar insights into better understanding worker behavior and assessing work quality \cite{rzeszotarski2011instrumenting,kazai2016quality} as well as detecting potential fraud \cite{heymann2011turkalytics}. Unfortunately, no open source project yet exists to enable other researchers to similarly instrument their own task interfaces. \turkey{} not only makes it easy to deploy external HITs on Mechanical Turk, but critically enables its users to capture and log underlying worker interactions on the front-end, presenting the potential to improve data quality and better understand latent worker behaviors underlying observed work products (e.g., behavioral trace data characterizing top performers).  

\turkey{} endeavors to be easy to use and expand. While \turkey{} itself is built atop the popular Python web framework Django, adding to its capabilities requires no more than a basic understanding of Python and JavaScript. \turkey{} seeks to provide requesters the freedom and ability to create their own components, as well as the option to reuse the foundational work of others. The benefit of a common, modular platform is that when a requester creates a new component to fill some gap, others should be able to understand and use it with relative ease. Preserving data export, admin creation process, and structure inherited from the framework which which users already have familiarity eases reusability and maintenance.

\begin{figure}[!htb]
\centering
\includegraphics[width=1\linewidth]{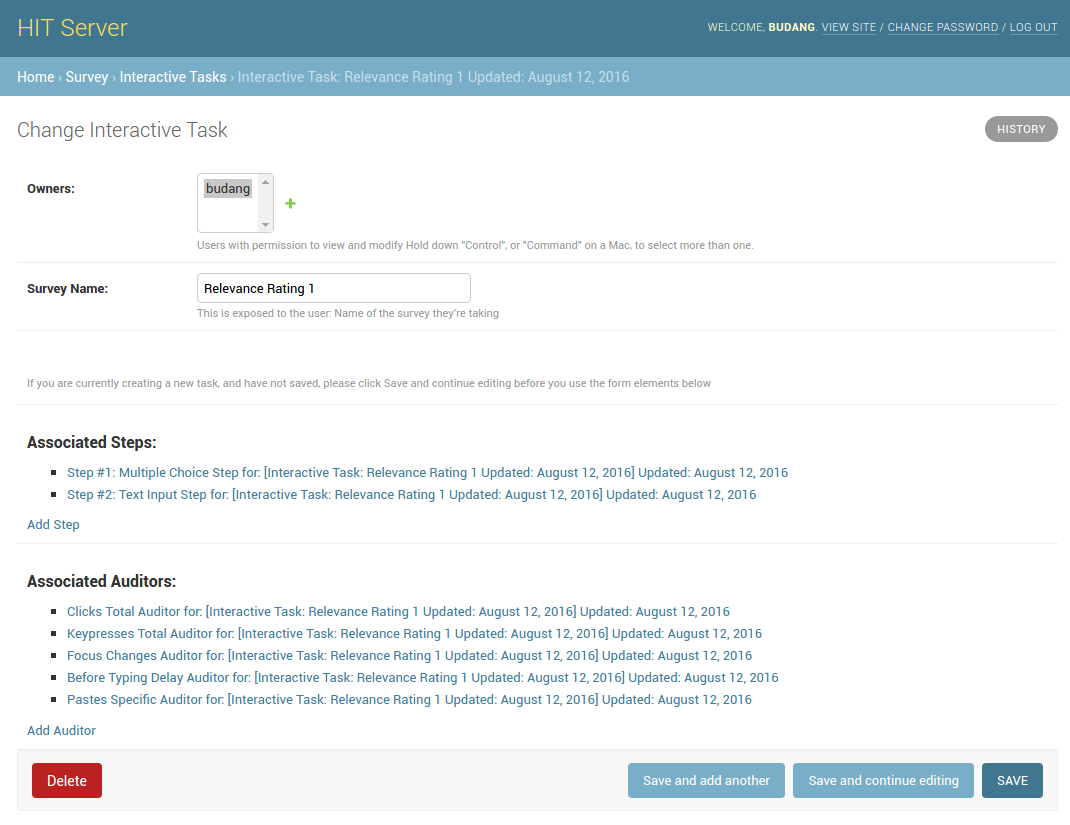}
\caption{\textit{Creating a new task}}
\label{fig:new_task}
\end{figure}

Components of the open source framework\footnote{\url{http://www.github.com/CuriousG102/turkey}} consist of: 
\begin{itemize}
\item \texttt{Task}s are HITs to be completed by workers. A task is comprised of steps and auditors and can be completed by any number of workers. Data will be recorded for each response. Figure \ref{fig:new_task} exemplifies the process of creating a new task in the online dashboard.
\item \texttt{Step}s are parts of a task that workers must complete and are the equivalent of a question. Steps are modular and can be assembled in any order, including a randomized order, within a task by the requester. Multiple choice, multiple answer (checkbox), and textual response types of steps have been implemented and are provided for immediate use. Requesters have the ability to create and add their own custom steps to a task if needed.
\item \texttt{Auditor}s, implemented in JavaScript and jQuery, surveil worker activity while completing a task and record user interaction with the webpage and browser. Many auditors are provided by default, including auditors that record the user's mouse movements, clicking interactions, and tab focus changes. Like steps, auditors are modular and requesters can choose which auditors to include in a task when creating one. Requesters may also create and add their own custom auditors as well.
\end {itemize}

For users satisfied with \turkey{}'s out-of-the-box capabilities, no extra code is necessary. Tasks are created within a browser-based interface in which authorized users can selectively add steps, auditors, and otherwise manage these tasks (Figure \ref{fig:new_task}). Data collected from a task is readily available to be exported as XML, which can be easily parsed to extract worker responses and auditor data (Figure \ref{fig:data_export}).

For those wanting to add their own steps or auditors, the following details of how the framework's administration, database, and export functionalities have been abstracted from the core information requirements of components:

\begin{itemize}
  \item What fields the new step or auditor will return when a user submits a response
  \item What fields the new step or auditor needs to pass to a user-provided template
  \item How to render a step on screen
  \item How to collect the information for the fields of an auditor
  \item Where to locate the JavaScript and HTML template that load alongside tasks
\end{itemize}

The administrative interface for the component, its database representation, and the logic to produce its XML export are all handled by \turkey{} (unless a user decides to override their hooks), reducing the time and complexity of adding custom items to the framework.

\begin{figure}[!htb]
{\scriptsize
\hphantom{}\texttt{$<$auditors$>$}\newline
\hphantom{12}\texttt{$<$clicks\char`_total$>$}\newline
\hphantom{1234}\texttt{$<$list\char`_item$>$}\newline
\hphantom{123456}\texttt{$<$model$>$survey.auditorclickstotaldata$<$/model$>$}\newline
\hphantom{123456}\texttt{$<$pk$>$1$<$/pk$>$}\newline
\hphantom{123456}\texttt{$<$fields$>$}\newline
\hphantom{12345678}\texttt{$<$general\char`_model$>$1$<$/general\char`_model$>$}\newline
\hphantom{12345678}\texttt{$<$count$>$4$<$/count$>$}\newline
\hphantom{123456}\texttt{$<$/fields$>$}\newline
\hphantom{1234}\texttt{$<$/list\char`_item$>$}\newline
\hphantom{12}\texttt{$<$/clicks\char`_total$>$}\newline
\hphantom{12}\texttt{...}\newline
\hphantom{}\texttt{$<$/auditors$>$}\newline
\caption{\textit{Example auditor data in exported XML}}
\label{fig:data_export}
}
\end{figure}


\section{Related Work}
\label{sec:related}

{\bf Task Fingerprinting}. \turkey{}'s auditors are based on the concept of \textit{task fingerprinting} \cite{rzeszotarski2011instrumenting}. Task fingerprinting is an attempt to capture the process in which workers work on and complete a specific task by using user event loggers. Information is recorded when a worker clicks, presses a key, or otherwise interacts with the webpage or browser. The collected data can then be used to analyze behavioral trends in "good" compared to "bad" workers and filter out potential low quality responses. 

{\bf Gold Standard Behavior}. Using task fingerprinting and a supervised classifier, an experiment was conducted to try and detect poor-performing workers based on their behavioral data as well as that of trained professional judges \cite{kazai2016quality}. Both normal workers and judges were given the same three tasks to complete, and 160 behavioral features were recorded per worker including "dwell" times between specific events (e.g., mouse clicks) and number of window resize events. Tasks were completed on \url{www.clickworker.com} for normal workers and on an in-house platform for judges. The recorded behavior of the judges was used as the gold standard.

{\bf TurkPrime} (\url{www.turkprime.com}) assists researchers in collecting data, particularly in the social and behavioral sciences \cite{litman2016turkprime}. Like \turkey{}, TurkPrime is integrated with Mechanical Turk and enables users to easily create and manage external HITs through an online administrative dashboard. TurkPrime offers many useful features, including the ability to create worker groups, include or exclude certain workers, and more options and control over managing HITs. Unlike \turkey{}, however, TurkPrime does not provide any form of auditor functionality. 

\nocite{atterer2006knowing}
\nocite{bakshy2014designing}
\nocite{hong2001webquilt}
\nocite{little2010turkit}
\nocite{nebeling2013crowdstudy}
\nocite{nebeling2013w3touch} \nocite{parkes2012turkserver}  

\section{Conclusion}
\label{sec:consclusion}
At its simplest, \turkey{} is a tool for easily developing and managing external HITs on Amazon Mechanical Turk. Though are other tookits have been created that offer some similar services, \turkey{} stands out as the first open source framework we are aware of providing auditors: a unique and powerful feature. Rather than just collect responses to tasks, these auditors can record a worker's interactions on the front-end, providing researchers and task designers of wealth of new data to study and better understand worker behaviors in task execution. \turkey{}, with its modular architecture and core auditor feature, enables its users to collect comprehensive data without the trouble of having to code a HIT that has already been created before.


{\bf Acknowledgments.} This work was made possible by the National Science Foundation's support for Research Experiences for Undergraduates (REU), under grant No. 1253413. The statements made herein are solely the responsibility of the authors. We thank our awesome crowd workers for their participation in powering our crowd-driven systems. 

\bibliographystyle{aaai}
\bibliography{bib}

\end{document}